\documentclass[11pt,a4paper]{article}

\usepackage[T1]{fontenc}
\usepackage[utf8]{inputenc}
\usepackage{lmodern}
\usepackage[a4paper,margin=2.6cm]{geometry}
\usepackage{amsmath,amssymb,bm}
\usepackage{graphicx}
\usepackage{xcolor}
\usepackage{microtype}
\usepackage{authblk}
\usepackage[numbers,sort&compress]{natbib}
\usepackage[hidelinks]{hyperref}

\title{Hidden-charm $udsc\bar c$ pentaquarks as flavor eigenstates in a constituent quark model}

\author[1,2]{M. C. Gordillo}
\author[1]{J. M. Alcaraz-Pelegrina}
\author[1]{J. Segovia\thanks{Corresponding author: 

\href{mailto:jsegovia@upo.es}{jsegovia@upo.es}}}

\affil[1]{Departamento de Sistemas F\'\i sicos, Qu\'\i micos y Naturales, Universidad Pablo de Olavide, Avenida Rectora Rosario Valpuesta 1, 41089 Dos Hermanas, Sevilla, Spain}

\affil[2]{Instituto Carlos I de F\'isica Te\'orica y Computacional, Universidad de Granada, E-18071 Granada, Spain}

\date{\today}

\begin{document}
\maketitle
\begin{abstract}
We use a diffusion Monte Carlo (DMC) algorithm to solve the Schr\"odinger equation describing $udsc\bar c$ pentaquarks within a non-relativistic constituent quark model. We consider multiquark states with well-defined parity, color, spin and isospin quantum numbers, consistent with the experimentally favored assignment $J^P = 1/2^-$ for one of the observed candidates, and impose $I = 0$. We find that, in order to reproduce the presence of multiple structures in the experimental mass region, the total wavefunction must also be an eigenstate of the SU(3) flavor operator. Under this condition, two structures emerge with masses lying in the same region as the experimentally observed signals in the $J/\psi\,\Lambda$ spectrum. In addition, two states are predicted below the $J/\psi\,\Lambda$ threshold but above the $\eta_c\,\Lambda$ one, which would not contribute to that channel. If only the $I = 0$ condition is imposed, a single structure is obtained, with a mass below the $J/\psi\,\Lambda$ threshold.
\end{abstract}


\section{Introduction}
\label{sec:introduction}

The first successful scheme to classify hadrons by their quark and antiquark numbers was proposed independently by Gell-Mann~\cite{Gell-Mann:1964ewy} and Zweig~\cite{Zweig:1964CERN}. This framework allows for configurations beyond the conventional $q\bar q$ and $qqq$ assignments corresponding to mesons and baryons, although clear experimental evidence for such exotic states only emerged several decades later. Reviews about those non-conventional structures, both from the experimental and theoretical points of view, can be found in Refs.~\cite{Chen:2016qju, Guo:2017jvc, Liu:2019zoy, Dong:2020hxe, Lebed:2023vnd, Hanhart:2025bun, Chen:2016spr, Yang:2020atz, Dong:2021bvy, Chen:2021erj, Mai:2022eur, Chen:2022asf, Ortega:2020tng, Huang:2023jec, Zou:2021sha, Du:2021fmf, Liu:2024uxn, Johnson:2024omq, Wang:2025dur, Wang:2025sic, Francis:2024fwf, Chen:2024eaq, Husken:2024rdk}.

In this context, the LHCb Collaboration observed two hidden-charm pentaquark candidates, with a minimal quark content $uudc\bar c$, in the $J/\psi\,p$ invariant mass spectrum of the decay $\Lambda_b^0 \to J/\psi\, K^-\, p$~\cite{LHCb:2015yax, LHCb:2019kea}. In 2020, evidence for an additional  structure in the $J/\psi \Lambda$ invariant mass spectrum in the decay $\Xi_b^- \to J/\psi\,\Lambda\,K^-$ was reported~\cite{LHCb:2020jpq}. This state, denoted $P_{cs}(4459)$, was identified as another hidden-charm pentaquark candidate, this time with a minimal quark content $udsc\bar c$. Its measured mass and width were:
\begin{align}
M &= 4458.8 \pm 2.9^{+4.7}_{-1.1}\,\text{MeV} \,, \\
\Gamma &= 17.3 \pm 6.5^{+8.0}_{-5.7}\,\text{MeV} \,.
\end{align}
with no spin-parity assignment. More recently, in 2022, another structure, $P_{cs}(4338)$, was observed with high statistical significance in the decay $B^- \to J/\psi\,\Lambda\,\bar p$~\cite{LHCb:2022ogu}. The reported mass and width are:
\begin{align}
M &= 4338.2 \pm 0.7 \pm 0.4\,\text{MeV} \,, \\
\Gamma &= 7.0 \pm 1.2 \pm 1.3\,\text{MeV} \,.
\end{align}
with a preferred assignment  $J^P = \tfrac{1}{2}^-$ ~\cite{ParticleDataGroup:2024cfk}.

Those experimental results have motivated a large amount of theoretical work. For instance, those systems have been studied using effective field theories~\cite{Feijoo:2022rxf, Zhu:2022wpi, Yan:2022wuz, Chen:2022onm, Zhu:2021lhd}, QCD sum rules~\cite{Wang:2022neq, Wang:2022gfb, Wang:2020eep, Chen:2020uif, Azizi:2023iym} or phenomenological quark models~\cite{Ortega:2022uyu, Giachino:2022pws, Yang:2022ezl, Wang:2022mxy, Karliner:2022erb, Meng:2022wgl, Shi:2021wyt, Xiao:2021rgp}. Many interpretations consider these states as hadronic molecules near the $\Xi_c \bar{D}^{(*)}$  and $\Xi^{(',*)}_c \bar{D}^{(*)}$ thresholds,  but other  configurations have also been proposed~\cite{Chen:2020kco, Chen:2022wkh, Li:2023aui, Maiani:2023nwj, Burns:2022uha}. 

In this work we use the diffusion Monte Carlo (DMC) technique to solve the Schr\"odinger equation that describes $udsc\bar c$ pentaquarks in the framework of a non-relativistic constituent quark model.  This method allows us to obtain the ground state of a given system, within statistical uncertainties,  providing that the initial approximation introduced in the algorithm (the so-called trial function) is reasonably close to the  exact ground-state wavefunction of the multiquark. In particular, the DMC is able to correct the shortcomings of that trial function if it has the same  nodal surface as the exact solution~\cite{hammond}. 

The Hamiltonian in the present study uses the AL1 potential defined in Refs.~\cite{Semay:1994ht, Silvestre-Brac:1996myf}, which includes a Coulomb term to model one-gluon exchanges, a linear confining potential, and a hyperfine spin-spin interaction. The parameters of the model were obtained from fits to the known properties of different baryons and were not changed here to describe larger multiquarks. The combination of DMC+ AL1 potential has been used successfully in previous works~\cite{Gordillo:2020sgc, Gordillo:2021bra, Alcaraz-Pelegrina:2022fsi, Gordillo:2022nnj, Gordillo:2023tnz, Gordillo:2024sem, Gordillo:2024blx, Gordillo:2025caj}. Given that the typical accuracy of the model in the conventional hadron sector is at the level of $(10-20)\%$, this range provides a reasonable estimate of the theoretical uncertainty of our predictions. The trial function we used contains a combination of the spin, color and flavor eigenvectors of the corresponding operators without any a priori couplings between baryon and meson parts. We also impose a definite value of the isospin for the flavor eigenfunctions \cite{Gordillo:2025caj}, since the proper consideration of the flavor degree of freedom is necessary to obtain the two pentaquark species found in the experiments (see below). 

Considering all that, the manuscript is organized as follows: in Sec.~\ref{sec:theory} we summarize the constituent quark Hamiltonian and briefly outline the DMC formalism. Section~\ref{sec:results} presents our  numerical results and discusses the mass spectrum and internal structure of the $udsc\bar c$ configurations. Finally, Sec.~\ref{sec:conclusions} contains our conclusions.

\section{Formalism}
\label{sec:theory}

We choose to describe the $udsc\bar c$ pentaquarks by the non-relativistic constituent quark model Hamiltonian: 
\begin{equation}
H = \sum_{i=1}^{5} \left( m_i + \frac{\vec{p\,}_i^{2}}{2m_i} \right) - \frac{\vec{P\,}_{\text cm}^{2}}{2M_{\text{tot}}} + \sum_{i<j} V_{ij} \,,
\end{equation}
where $m_i$ are the constituent quark masses, $P_{\text cm}$ is the total center-of-mass momentum, and $V_{ij}$ denotes the pairwise interaction between the $i$ and $j$ quarks. That interquark interaction was taken to be the standard AL1 potential defined in Refs.~\cite{Semay:1994ht, Silvestre-Brac:1996myf}:
\begin{align}
V_{ij} = (\vec{\lambda}_i^c\cdot\vec{\lambda}_j^c) \Bigg[ \frac{\kappa}{r_{ij}} - \lambda r_{ij} + \Lambda - \frac{2\pi}{3m_im_j} \kappa^\prime (\vec{\sigma}_i\cdot\vec{\sigma}_j) \frac{e^{-r_{ij}^2/r_0^2}}{\pi^{3/2}r_0^3} \Bigg] \,,
\end{align}
where $\vec{\lambda}^c$ are the SU(3) color Gell-Mann matrices and $\vec{\sigma}$ are the Pauli matrices. The regulator is defined as
\begin{equation}
r_{0}(m_i,m_j) = A \left(\frac{2m_im_j}{m_i+m_j}\right)^{-B} \,,
\end{equation}
with quark masses 
\begin{align}
m_u=m_d &= 0.315\,\text{GeV} \,, \nonumber \\
m_s &= 0.577\,\text{GeV} \,, \nonumber \\
m_c &= 1.836\,\text{GeV} \,, \nonumber
\end{align}
and the remaining parameters defined as
\begin{eqnarray}
\kappa = 0.0951\,, & \quad\quad & \kappa^\prime = 0.3489 \,, \nonumber \\
\lambda = 0.0310\,\text{GeV}^2 \,, & \quad\quad & \Lambda = 0.1560\,\text{GeV} \,,  \nonumber \\
B = 0.2204 \,, & \quad\quad & A = 1.6553\,\text{GeV}^{B-1} \,. \nonumber
\end{eqnarray}

The diffusion Monte Carlo (DMC) technique was developed in the context of condensed matter physics to solve many-body Schr\"odinger equations. This makes it particularly well suited to deal with large ensembles of quarks~\cite{hammond}. In the DMC method, we start by writing the Schr\"odinger equation in imaginary time ($\hbar=c=1$) as:
\begin{equation}
-\frac{\partial \Psi (\bm{R},\alpha,t)}{\partial t}
= (H-E_s)\,\Psi(\bm{R},\alpha,t)\,,
\label{eq:Sch1}
\end{equation}
where $E_s$ is an appropriate energy shift and $\bm{R}\equiv(\vec{r}_1,\ldots,\vec{r}_n)$ stands for the positions of the $n$ quarks and $\alpha$ denotes each possible spin-flavor-color state (channel in the standard literature), with given quantum numbers. Then, we expand $\Psi(\bm{R},\alpha,t)$ in terms of the complete set of the Hamiltonian's eigenfunctions as
\begin{equation}
\Psi(\bm{R},\alpha,t) = \sum_{i,\alpha} c_{i,\alpha}\, e^{-(E_i-E_s)t} \, \phi_i(\bm{R},\alpha) \,,
\end{equation}
where the $E_i$ are the eigenvalues of the system's Hamiltonian operator and the $\phi_i(\bm{R},\alpha)$ their corresponding eigenfunctions.  The ground state wave function is obtained as the asymptotic solution of 
Eq.~\eqref{eq:Sch1} when $t\to \infty$.  

From all this, we can develop an operational recipe to apply the DMC algorithm with importance sampling to a given system \cite{Gordillo:2020sgc}.  This involves the use of an initial approximation to the ground state of the system,  the so-called trial function, $\Psi_{T} (\bm{R}, \alpha)$, that includes all the information known \emph{a priori} about the system. In our approach,  the trial wave function is constructed as a single state in the full Hilbert space,
\begin{align} \label{eq:Eigenfunction}
\Psi_{T}(\bm{R},\alpha)=\phi_T(\bm{R})\sum_\alpha d_\alpha\chi^{(\alpha)} \,,
\end{align}
where $\phi_T(\bm{R})$ depends on the quark coordinates and the coefficients $d_\alpha$ encode the spin($s_i$)--color($c_i$)--flavor($f_i$) structure of the spin-color-flavor function $\chi^{(\alpha)} (\{s_i\}, \{f_i\}, \{c_i\})$ corresponding to the $\alpha$ state, for  every $i$ quark. In the DMC algorithm with importance sampling,  the evolution of the system is guided not by $[\Psi_{T}(\bm{R},\alpha)]^2$ as in a purely variational method,  but by the product
\begin{equation}
f(\bm{R},\alpha,t) = \Psi(\bm{R},\alpha,t)\Psi_{T}(\bm{R},\alpha) \,.
\end{equation}
This $f(\bm{R},\alpha, t)$ is represented by a set of {\em walkers} that include both the  particle coordinates {\em and} the coefficients of the spin-color-flavor functions~\cite{Gordillo:2020sgc}. In a DMC step,  both the coordinates and coefficients are sampled at the same time, since the  Hamiltonian couples all channels. As a consequence, the resulting state cannot be interpreted as a set of independent channels to which we could assign separate spatial wavefunctions. The robustness of this approximation was tested in Ref.~\cite{Gordillo:2020sgc}, where the masses of tetraquarks obtained compared favorably with the ones obtained by variational calculations using the same potential. 

In Eq.~\eqref{eq:Eigenfunction}, the functions $\chi^{(\alpha)}$ are built from the eigenfunctions of the spin, {\em flavor} and color operators, given by: 
\begin{align}
S^2 &= \left(\sum_{i=1}^{5} \frac{\vec{\sigma}_i}{2}\right)^2 \,
F^2 = \left(\sum_{i=1}^{n_f} \frac{\vec{\lambda}_i^f}{2}\right)^2 \,
C^2 = \left(\sum_{i=1}^{5} \frac{\vec{\lambda}_i^c}{2}\right)^2 \,,
\label{eq:operators_penta}
\end{align}
with well-defined eigenvalues. $\vec{\lambda}$ are the standard Gell-Mann matrices acting in the color space, while $\vec{\lambda}_i^f$ are the same Gell-Mann matrices but this time acting in flavor space. The flavor operator acts only on the three light quarks $u,d,s$ ($n_f=3$); the charm quark and antiquark being excluded from the SU(3) flavor classification. The Pauli matrices are $\vec{\sigma}$. When constructing the proper spin-color-flavor functions, the $u$, $d$ and $s$ quarks are {\em always} indistinguishable.

In this work we restrict ourselves to total orbital angular momentum $L^2=0$, corresponding to the $J^P=1/2^{-}$ state assigned experimentally to the $P_{cs}(4338)$ pentaquark~\cite{LHCb:2022ogu}. Under this assumption, the spatial part of the trial wave function should depend only on the set of inter-particle distances, which makes it symmetric under permutations of the spatial coordinates. To fulfill this requirement, we adopt the Jastrow-type ansatz,
\begin{equation}
\phi_T (\vec{r}_1,\ldots,\vec{r}_n) = \prod_{j>i=1}^{n} \exp(-a_{ij} r_{ij}) \,,
\label{eq:radialwf}
\end{equation}
where the coefficients $a_{ij}$ are chosen so as to satisfy the cusp conditions derived from the Coulomb-like part of the interquark interaction for each pair of constituents. This is the function compatible with the $L^2=0$ construction with the lowest number of nodes, that it should produce the lowest masses~\cite{hammond} for the pentaquarks. The spin sector is constructed by coupling the spins of the five quarks to the desired total spin $S=1/2$  to produce the 5 possible (degenerate) spin wavefunctions. On the other hand, the total wavefunction has to be colorless. For a $5$-quark system, there are 3 linearly independent such degenerate configurations.

Special care must be taken when defining the flavor wave function. Both $udsc\bar c$ pentaquark candidates, $P_{cs}(4338)$ and $P_{cs}(4459)$, have been observed as enhancements in the $J/\psi\,\Lambda$ invariant-mass distribution. This implies that both pentaquarks should have isospin $I=0$.   The three eigenfunctions of the isospin operator with eigenvalue equal to zero are: 
\begin{align}
\chi_{f1} &= \frac{1}{\sqrt{2}} (sud - sdu) \,, \nonumber \\
\chi_{f2} &= \frac{1}{\sqrt{2}} (uds - dus) \,, \nonumber \\
\chi_{f3} &= \frac{1}{\sqrt{2}} (usd - dsu) \,. \nonumber
\end{align}
None of these functions are independently eigenvectors of the flavor operator.  On the other hand, we can combine them to produce two sets of independent eigenfunctions of that operator, one singlet with eigenvalue $F^2= 0$:
\begin{equation}
\chi_f^{(1)} \;=\; \frac{1}{\sqrt{6}}\Big( uds - dus + dsu - sdu + sud - usd \Big)\,.
\end{equation}
that can be expressed as the linear combination:
\begin{equation}
\chi_f^{(1)}
= \frac{1}{\sqrt{3}}\left(\chi_{f1}+\chi_{f2}-\chi_{f3}\right)\,,
\end{equation}
and two of the functions of the octet manifold with $F^2$ =3;
\begin{equation}
\chi_{f,1}^{(8)} \;=\; \frac{1}{2\sqrt{3}}\Big[(sd-ds)u + (us-su)d + 2(ud-du)s\Big]\,,
\end{equation}
\begin{equation}
\chi_{f,2}^{(8)} \;=\; \frac{1}{2}\Big[(sd+ds)u - (su+us)d\Big]\,.
\end{equation}
that can be written as (respectively):
\begin{equation}
\chi_{f,1}^{(8)}
= \frac{1}{\sqrt{6}}\left(-\chi_{f1}+2\,\chi_{f2}+\chi_{f3}\right)\,,
\end{equation}
\begin{equation}
\chi_{f,2}^{(8)}
= -\frac{1}{\sqrt{2}}\left(\chi_{f1}+\chi_{f3}\right)\,.
\end{equation}
Those two functions are the same as those corresponding to the $\Lambda$ baryon~\cite{libro2018}, and have,  by construction an isospin $I=0$. Taking everything into account, the pentaquark system allows for two independent families of states with definite quantum numbers:
\begin{align}
3 \, \text{(colorless)} \times 5 \, (S=1/2) \times 1 \, (F^2= 0,I=0) &= 15 \, \text{functions} \,, \nonumber \\
3 \, \text{(colorless)} \times 5 \, (S=1/2) \times 2 \, (F^2=3,I=0)  &= 30 \, \text{functions} \,, \nonumber
\end{align}
from which we have to produce linear combinations that will have to be multiplied by the  spatial part given by Eq.~\eqref{eq:radialwf} to build the total trial function defined by Eq.~\eqref{eq:Eigenfunction}. Those linear spin-color-flavor combinations are made of terms of the type (for instance):
\begin{align}\label{sample}
|\uparrow ru, \uparrow g s,\downarrow b d,  \uparrow r,
\downarrow \bar{r}\rangle \,, \nonumber \\
|\uparrow rs, \uparrow gd,\downarrow gu,  \uparrow b,
\downarrow \bar{g}\rangle \,,
\end{align}
(no flavor for $c$ and $\bar{c}$, that are always located in the 4th and 5th positions).  

However, not all these functions are admissible, since antisymmetry must be imposed with respect to the exchange of identical quarks, since quarks are fermions. A wavefunction is antisymmetric when it is an eigenvector of the operator:
 \begin{equation}
\mathcal{A} = \frac{1}{N_p} \sum_{{\alpha}=1}^{N_p} (-1)^P \mathcal{P_{\alpha}} \,,
\label{eq:antisymope}
\end{equation}
applied only to the spin-color-flavor part of wavefunction, since the radial part is symmetrical with respect to the exchange of any quarks.  In Eq.~\eqref{eq:antisymope}, $N_p$ is the number of possible permutations of the quark indexes, $P$ is the order of the permutation, and  $\mathcal{P_{\alpha}}$'s  stand for the matrices that define those permutations. Once the matrix derived from the operator in Eq.~\eqref{eq:antisymope} is constructed, using as a basis comprising one of the sets of spin-color-flavor functions defined above (15 or 30 functions, respectively),  we obtain its eigenvectors as linear combinations of those (15 or 30) functions. The eigenvalues of the corresponding matrix can be only  0 or 1.   Only the combinations with an eigenvalue equal to one are antisymmetric with respect to the exchange of quarks, i.e., they correspond to the  $\chi^{(\alpha)}$'s in Eq.~\eqref{eq:Eigenfunction}.  However  if, instead of having one single function compatible with all restrictions, we have two or more,  there is an additional complication: any rotation of that set of  functions in the Hilbert space is also a valid input in the DMC algorithm since it will have the same set of spin, color, isospin and flavor quantum numbers \cite{Gordillo:2025caj}. Fortunately, physical observables depend only on the subspace spanned by these functions, not on a particular basis choice. 

To completely define the $\chi^{(\alpha)}$ functions, we recall that the antisymmetry condition must be imposed only on {\em identical} quarks. For instance, in the operator defined by Eq.~\eqref{eq:antisymope}, the permutation operators $\mathcal{P}_{\alpha}$ cannot exchange the full set of quantum numbers (spin, color and flavor) of quarks $3$ and $4$, since, according to our convention (see Eq.~\eqref{sample}), the quark in the fourth position is a $c$ quark and is therefore distinguishable from any of the light quarks. Within this framework, and following the strategy adopted in our previous work~\cite{Gordillo:2024sem}, we consider different symmetry-constrained constructions of the spin-color-flavor wave function, corresponding to different patterns of internal correlations among the quarks.  For instance,  a hidden-charm hexaquark can be described by either a baryon-antibaryon correlated structure, or a combination of a tetraquark-meson-like arrangement, in both cases including hidden color contributions to the description of the multiquark.  Or a hidden-charm pentaquark can be modeled as a baryon-meson-like structure or a diquark-diquark-antiquark one \cite{Gordillo:2024sem}. These constructions do not necessarily define orthogonal sectors of the full Hilbert space, but different internal functions that can be statistically separated using several measurements but not in a single experiment as in the case of orthogonal states~\cite{TongTopicsQM}.  In particular, we analyze the following two set of functions:

\begin{itemize}

\item A first construction in which the three light quarks are treated on equal footing and full antisymmetry under their exchange is imposed. This realization naturally accommodates configurations reminiscent of a $qqq$--$c\bar{c}$ structure. However, it does {\em not} correspond to a direct product of a $uds$ baryon and a $c\bar{c}$ meson. Hidden-color components are included (see, for instance, the lower term in Eq.~\eqref{sample}), and no assumption about spatial separation is made, since this can only be inferred from the distributions obtained in the DMC calculation. In this construction, configurations analogous to $\Lambda + c\bar{c}$ can be represented, whereas $\Xi_c + \bar{D}$-like arrangements are not naturally accommodated, because the two light quarks in the $qqc$ cluster are not equivalent to the light quark associated with the $\bar{c}$.

This construction leads to 2 antisymmetric functions with $F^2=0$ and 5 antisymmetric functions with $F^2=3$. The first (second) set of 2 (5) functions corresponds to linear combinations of the 15 (30) spin-color-flavor basis states with the appropriate quantum numbers defined above. Each set defines a pentaquark state, since the corresponding channels enter Eq.~\eqref{eq:Eigenfunction}. We denote these states as Ia (2 channels) and Ib (5 channels), respectively.

\item A second construction in which the trial wave function is organized so as to emphasize configurations where two light quarks correlate with the $c$ quark, forming a $qqc$ structure, while the remaining light quark correlates with the $\bar{c}$ antiquark to form a $q'\bar{c}$ component. This leads to a $qqc$-$q'\bar{c}$ realization in which the first two light quarks are treated equally, but not on the same footing as the third one. Accordingly, in Eq.~\eqref{eq:antisymope}, the permutation operators $\mathcal{P}_{\alpha}$ act only on the first two quarks. As in the previous case, this construction does {\em not} imply a baryon-meson decomposition, since hidden-color configurations are included and no spatial separation is assumed. It does not single out any particular flavor either, as the configurations $usc+d\bar{c}$, $dsc+u\bar{c}$ and $udc+s\bar{c}$ are treated on the same footing and enter the same linear combinations defining the $\chi^{(\alpha)}$. Within this realization, configurations analogous to $\Lambda + c\bar{c}$ are not accessible.

This construction yields 7 antisymmetric functions with $F^2=0$ and 15 antisymmetric functions with $F^2=3$. These combinations define two additional pentaquark states, denoted IIb and IIa, respectively (see discussion below).
\end{itemize}

We stress that this separation in the spin-color-flavor space reflects different internal wave functions and does not necessarily correspond to any physical or spatial separation of the quarks. This is explicitly confirmed by the spatial distributions obtained in the DMC calculation.

Having completed this construction,  we can apply the DMC algorithm to the full trial function (in fact four different trial functions) defined in Eq.~\eqref{eq:Eigenfunction}. In our implementation, the walkers carry fixed spatial labels for the quarks, while the spin-color-flavor part of the wavefunction is written as a linear combination of terms involving different flavor permutations of the light quarks. As a consequence, for a given spatial label, the quark masses are not uniquely defined term by term. This is handled by evaluating the  Hamiltonian at each DMC step using the instantaneous internal wavefunction corresponding to each walker. In particular, the quantities that depend on the quark masses, such as the kinetic and potential terms,  are calculated as weighted values for each internal state defined by the $d_{\alpha}$'s. We have verified that this procedure is numerically stable and that the final results are insensitive, within statistical uncertainties, to both the choice of the coordinate part of the radial function and the initial set of walkers (including coordinates and $d_{\alpha}$'s)~\cite{Gordillo:2025caj}.




\section{Results}
\label{sec:results}

\subsection{Masses}

In this section, we present our numerical results of the mass spectrum and internal structure of the possible $J^P=1/2^-$, $I=0$ $udsc\bar c$ pentaquarks. The obtained masses are as follows,
\begin{align} \nonumber
M(\mathrm{Ia},F^2=0)  &= 4473 \pm 5~\text{MeV} \,, \\ \nonumber
M(\mathrm{Ib},F^2=3)  &= 4237 \pm 3~\text{MeV} \,, \\[2ex] \nonumber
M(\mathrm{IIa},F^2=3) &= 4350 \pm 6~\text{MeV} \,, \\ \nonumber
M(\mathrm{IIb},F^2=0) &= 4245 \pm 7~\text{MeV} \,. \nonumber
\end{align}
where the $a$ subscripts were assigned to the pentaquarks with the larger masses. For comparison, the relevant baryon-meson thresholds obtained with the same AL1 Hamiltonian using DMC are:
\begin{align} \nonumber
\eta_c + \Lambda &: 4161~\text{MeV} \,, \\ \nonumber
J/\psi + \Lambda &: 4257~\text{MeV} \,,
\end{align}
which have been determined by calculating within the same theoretical framework the masses of the light $(uds)$-baryon with $I=0$, $J=1/2^+$ ($\Lambda$-like configuration):
\begin{equation}  \nonumber
M(\Lambda) = 1156 \pm 2~\text{MeV} \,,
\end{equation}
to be compared with $M(\Lambda)=1154$ MeV of Ref.~\cite{Silvestre-Brac:1996myf}. For the lowest $s$-wave charmonium states, we have~\cite{Gordillo:2020sgc}. 
\begin{align} \nonumber
M(\eta_c) &= 3005~\text{MeV} \,, \nonumber \\
M(J/\psi) &= 3101~\text{MeV} \,.
\end{align}
Therefore, pentaquark Ia lies above the $J/\psi\,\Lambda$ threshold and can decay into this experimentally observed channel. In contrast, pentaquark Ib lies below $J/\psi\,\Lambda$ but above $\eta_c\,\Lambda$, implying that it should predominantly decay by that channel. The same pattern holds for IIa and IIb.

Within this approach, two structures emerge in the mass region of the experimentally observed $P_{cs}$ states. While they cannot be interpreted as arising from independent Hilbert spaces, their presence reflects the richness of the correlation dynamics in the multiquark system, which may manifest experimentally as multiple nearby enhancements in the invariant-mass spectrum.

The states Ia and IIa have the same set of spin, color and isospin eigenvalues, but belong to different flavor states. The state Ia corresponds to $F^2=0$, while IIa belongs to the $F^2=3$ sector. This means that their spin-color-flavor wavefunctions probe different components of the full Hilbert space and suggest the presence of multiple structures associated with different dominant correlation patterns within the same multiquark system. The masses obtained for these states lie in the same energy region as the experimentally observed $P_{cs}$ signals, within the uncertainties of the model. 
A possible reason why the additional states (Ib and IIb) have not been observed so far is that the $\eta_c\,\Lambda$ decay channel, into which they are expected to decay, has not yet been experimentally investigated.

To reproduce the emergence of multiple structures in the experimental mass region, we have to consider eigenfunctions of the flavor (not only isospin) operator.  If we break the flavor symmetry at the level of the wavefunction, we end up with a single $I=0$ configuration:
\begin{align}
\chi_{f2} &= \frac{1}{\sqrt{2}} (uds - dus)\,, \nonumber
\end{align}
which leads to a single family of functions:
\begin{align*}
3 \, \text{(colorless)} \times 5 \, (S=1/2) \times 1 \, (I=0) &= 15 \, \text{functions} \,, \nonumber 
\end{align*}
that are not eigenvectors of the full flavor operator. After applying the antisymmetric operator, we are left with 7 functions (out of 15), whose combinations produce a single state (pentaquark III), whose quantum numbers are  also compatible with the experimental  assignment $J^P=1/2^-$ and $I=0$. We predict its  mass to be:
\begin{equation} \nonumber
M(\mathrm{III}) = 4200 \pm 4~\text{MeV} \,.
\end{equation}
This state lies below the $J/\psi\,\Lambda$ threshold and therefore cannot decay into this channel.  Consequently, it cannot be the one observed in the  experimental analyses. This indicates that a consistent treatment of the SU(3) flavor symmetry is necessary to reproduce multiple structures in the experimental mass region.


\begin{table}[!t]
\caption{\label{tab:rij2Ia} Average distances for $J^P=1/2^-$ $udsc\bar c$ Ia and Ib pentaquarks  in fm.  Read related text of Sec.~\ref{sec:theory} for notation of constituent quarks and pentaquarks. In all cases, the error bars are $\pm$ 0.02 fm.}
\vspace*{0.20cm}
\centering
\begin{tabular}{lcccc}
\hline
State & $r_{qq}$ & $r_{qc}$ & $r_{q\bar c}$ & $r_{c\bar c}$ \\
\hline
Ia & 0.80 & 0.70 & 0.67 & 0.64 \\
Ib & 0.78 & 1.55 & 1.55 & 0.33 \\
\hline
\end{tabular}
\end{table}

\begin{table}[!t]
\caption{\label{tab:rij2IIa} Mean interquark distances (in fm) for the
$J^P=1/2^-$ hidden-charm $uds c\bar{c}$ pentaquark configurations Ia and Ib. The notation follows the quark assignments introduced in Sec.~\ref{sec:theory}. The statistical uncertainties are $\pm 0.02\,\text{fm}$ for all reported values.}
\vspace*{0.20cm}
\centering
\begin{tabular}{lcccccc}
\hline
State & $r_{qq'}$ & $r_{qq}$ & $r_{qc}$ & $r_{q'\bar c}$ & $r_{q'c}$ & $r_{c\bar c}$ \\
\hline
IIa & 0.95 & 0.72 & 0.66 & 0.54 & 0.83 & 0.73 \\
IIb & 0.81 & 0.74 & 1.30 & 1.31 & 1.31 & 0.34 \\
\hline
\end{tabular}
\end{table}

\subsection{Internal structure}

To characterize the spatial structure of the pentaquarks that have been located in the same mass region as the experimentally observed $P_{cs}$ structures, we provide the mean interquark distances $\langle r_{ij} \rangle$ (in fm) in Tables~\ref{tab:rij2Ia} and~\ref{tab:rij2IIa}, respectively. We also give the same values for structures Ib and IIb for the sake of comparison. These results indicate that both Ia and IIa correspond to compact configurations, with all interquark separations lying in the range $\langle r_{ij} \rangle \sim 0.7$--$1.0$ fm, and no evidence of a large baryon--meson separation as expected in loosely bound molecular configurations. Since both states have a different set of quantum numbers, this means that the DMC algorithm is not determined exclusively by the internal wavefunction, but takes into account the full correlations defined by the Hamiltonian.

As indicated above, the comparison of these configurations is meaningful even though the different antisymmetrization prescriptions do not necessarily define mutually orthogonal sectors of the full Hilbert space. In general, non-identical, non-orthogonal wavefunctions can still be distinguished statistically through suitable observables, since they may yield different measurement probabilities or expectation values \cite{TongTopicsQM}. In the present case, the flavor-singlet ($F^2=0$) and flavor-octet ($F^2=3$) configurations are distinguished by their flavor quantum numbers. More generally, different internal correlation patterns may lead to distinct masses, interquark distances, spatial correlation functions, or overlaps with baryon--meson channels~\cite{Gordillo:2026usv, Gordillo:2026ztm}. The analysis of such quantities therefore provides means of characterizing different pentaquark configurations, without requiring their interpretation as states belonging to disjoint sectors of the full Hilbert space.

To go further in the characterization of the structures, we have compared those internal distances to those of the isolated mesons and baryons. The corresponding values (in fm) are:
\begin{align}  \nonumber
\langle r_{\bar c c} \rangle_{\eta_c} = 0.34 \,,\quad
\langle r_{\bar c c} \rangle_{J/\psi} &= 0.37\,,
\end{align}
for the $\bar c c$ meson,  while for the $\Lambda$ baryon we have $\langle r_{qq} \rangle_{\Lambda}$ = 0.80 fm,  with uncertainties $\pm$ 0.02 fm in all cases. Therefore, the Ia pentaquark resembles a distorted $\Lambda + c\bar c$ configuration, with a significantly enlarged $c\bar c$ separation relative to charmonium. On the other hand, for the IIa pentaquark,  the distances needed for the comparison are those corresponding to open-charm baryons and $\bar c q$ mesons. Those are (in fm):
\begin{align} \nonumber
\text{$udc$}: \quad &\langle r_{qq} \rangle = 0.76 \,, \quad \langle r_{qc} \rangle = 0.71, \,\\ \nonumber
\text{$usc$}: \quad & \langle r_{qq} \rangle= 0.71 \,, \quad \langle r_{qc} \rangle = 0.66, \,
\end{align}
with 
\begin{align} \nonumber
\langle r_{\bar c u} \rangle_{\bar D} = 0.59 \,, \quad 
\langle r_{\bar c s} \rangle_{\bar D_s}= 0.57 \,, \\ \nonumber
\langle r_{\bar c u} \rangle_{\bar D^*} = 0.67 \,, \quad 
\langle r_{\bar c s} \rangle_{\bar D^*_s}= 0.54 \,,  \nonumber
\end{align}
with the same $\pm$ 0.02 fm uncertainties. This suggests that the IIa pentaquark is structurally closer to a compact $qqc + q'\bar c$ arrangement with the baryon + meson very close to each other. On the other hand, the Ib and IIb structures resemble in both cases a baryon + $\bar c c$ meson whose constituents are relatively far apart. One can also see that the symmetry restrictions imposed to the wavefunctions that model pentaquarks Ia and Ib produce different internal structures.  


\section{Conclusions}
\label{sec:conclusions}

We have investigated the hidden-charm pentaquark system with quark content $udsc\bar c$ within a non-relativistic constituent quark model using the AL1 interaction, solving the corresponding Schr\"odinger equation by means of the diffusion Monte Carlo (DMC) technique. The calculation has been performed by constructing trial wavefunctions with well-defined quantum numbers in the spin, color and flavor (not only isospin) spaces, and imposing the appropriate antisymmetry constraints on the internal degrees of freedom.

By combining the flavor eigenvectors with the full set of colorless and $S = 1/2$ spin functions, we obtain four symmetry-constrained $J^{P} = 1/2^{-}$ $udsc\bar c$ configurations. Two of them lie above the $J/\psi\,\Lambda$ threshold, with masses $M_{\text{Ia}} = 4473 \pm 5$ MeV and $M_{\text{IIa}} = 4350 \pm 6$ MeV, and fall within the mass region where the experimentally observed $P_{cs}$ structures are located. The remaining two states lie below the $J/\psi\,\Lambda$ threshold but above the $\eta_c\,\Lambda$ one, suggesting that they would predominantly decay into the latter channel, which has not yet been experimentally explored. Those states are different, and hence distinguishable by performing several (not only one) set of measurements. 


If, instead of considering the full $I = 0$ flavor subspace, the strange quark is treated as distinguishable and the flavor sector is restricted to a single isospin eigenfunction, only one pentaquark state is obtained. Its predicted mass, $M_{\text{III}} = 4200 \pm 4$ MeV, lies below the $J/\psi\,\Lambda$ threshold and therefore cannot be associated with the observed structures. This indicates that a consistent treatment of SU(3) flavor symmetry is necessary to reproduce multiple structures in the experimental mass region.

%
%

\section*{Acknowledgments}

The authors gratefully acknowledge the computational resources provided by the C3UPO High Performance Computing Center at Universidad Pablo de Olavide, Seville, Spain.

\section*{Funding}
This work was supported by the Spanish Ministerio de Ciencia e Innovaci\'on under Grant Nos.\ PID2020-113565GB-C22 and PID2022-140440NB-C22, and by the Junta de Andaluc\'ia under Grant Nos.\ PAIDI FQM-205 and FQM-370.

\section*{Author contributions}

M.C. Gordillo: Conceptualization, Methodology,
Formal analysis, Software, Investigation, Writing -- original draft, Writing -- review \& editing, Funding acquisition.

J.M. Alcaraz-Pelegrina: Methodology, Formal analysis, Software, Validation, Writing -- review \& editing.

J. Segovia: Conceptualization, Formal analysis, Investigation, Writing -- original draft, Writing -- review \& editing, Funding acquisition.

All authors have read and approved the final version of the manuscript.

\section*{Data availability}
The data that support the findings of this article are not publicly available. The data are available from the authors upon reasonable request.

%


\bibliographystyle{unsrtnat}
\bibliography{arXiv_DMC-Tccs}

\end{document}